\documentclass{aa}

\usepackage {epsfig,graphicx,color}
\usepackage {txfonts   }
\usepackage {natbib    }
\usepackage {float     }
\usepackage {enumerate }
\usepackage {tabularx  }
\usepackage {url       }
\usepackage {capt-of   }
\usepackage {color     }

\begin{document}

%

\title{Age, size, and position of \ion{H}{ii} regions in the Galaxy}
\subtitle{Expansion of ionized gas in turbulent molecular
  clouds}

\author{ 
P. Tremblin    \inst{1,2,3}\and 
L. D. Anderson \inst{4}\and
P. Didelon     \inst{5}\and
A. C. Raga     \inst{6,3}\and
V. Minier      \inst{5}\and
E. Ntormousi   \inst{5}\and
A. Pettitt     \inst{1}\and
C. Pinto       \inst{7}\and
M. Samal       \inst{7}\and
N. Schneider   \inst{8,9}\and
A. Zavagno     \inst{7}
       }

\institute{
  Astrophysics Group, University of Exeter, EX4 4QL Exeter, UK
  \and
  Maison de la Simulation, CEA-CNRS-INRIA-UPS-UVSQ, USR 3441, Centre
  d'\'etude de Saclay, 91191 Gif-Sur-Yvette, France
  \and
  Nordita, KTH Royal Institute of Technology 
  and Stockholm University, Roslagstullsbacken 23, 10691 Stockholm, Sweden
  \and
  Department of Physics, West Virginia University, Morgantown, WV 26506, USA
  \and 
  Laboratoire AIM Paris-Saclay (CEA/Irfu - Uni. Paris Diderot
  - CNRS/INSU), Centre d'\'etudes de Saclay,  91191 Gif-Sur-Yvette,
  France
  \and
  Instituto de Ciencias Nucleares, Universidad Nacional Aut\'onoma de M\'exico, Ap. 70-543, 04510 DF, Mexico
  \and
  Aix Marseille Universit\'e, CNRS, LAM (Laboratoire
  d'Astrophysique de Marseille) UMR 7326, 13388, Marseille,
  France
  \and
  Univ. Bordeaux, LAB, UMR 5804, F-33270, Floirac, France
  \and
  CNRS, LAB, UMR 5804, F-33270, Floirac, France
}

\date{\today}
\mail{tremblin@astro.ex.ac.uk}

\titlerunning{\ion{H}{ii} regions in turbulent molecular clouds}
\authorrunning{Tremblin et al.}

\abstract{}
{This work aims at improving the current understanding of the
  interaction between \ion{H}{ii} regions and turbulent molecular
  clouds. We propose a new method to determine the age of a large sample of OB
  associations by investigating the development of their associated
  \ion{H}{ii} regions 
  in the surrounding turbulent medium. }
{Using analytical solutions, one-dimensional (1D), and
  three-dimensional (3D) simulations, we constrained the
  expansion of the ionized bubble depending on the turbulent level of
  the parent molecular cloud. A grid of 1D simulations was then
  computed in order to build isochrone curves for \ion{H}{ii} regions
  in a pressure-size diagram. This grid of models allowed to date large sample
of OB associations and was used on the \ion{H}{ii} Region Discovery
Survey (HRDS).}
{Analytical solutions and numerical simulations showed that the
  expansion of \ion{H}{ii} regions is slowed down by the turbulence
  up to the point where the pressure of the ionized gas is in a
  quasi-equilibrium with the turbulent ram pressure. Based on this
  result, we built a grid of 1D models of the expansion of \ion{H}{ii}
  regions in a profile based on Larson laws. The 3D turbulence is
taken into account by an effective 1D temperature profile. The ages
estimated by the isochrones of this grid agree well with literature
values of well-known regions such as Rosette,
RCW~36, RCW~79, and M16. We thus propose that this
method can be used to give ages of young OB associations through the
Galaxy such as the HRDS survey and also in nearby extra-galactic sources.} 
{}


\maketitle

%
%

\section{Introduction}

The age of a star cluster can be derived using photometry or
spectroscopy and
evolutionary tracks in the Hertzsprung-Russell (HR) diagram
\citep[e.g.][]{Meynet:2003kc,Martins:2010jw,Martins:2012gv}. In order
to be precise, photometric methods 
require the analysis of a large statistic of stars which can be tedious, and is
very difficult 
for a single object because of the uncertainties. Spectroscopic
methods are more reliable but the analysis of a lot of objects is also
needed to derive an estimation of the cluster age. When the star
cluster contains massive stars that 
ionize their environment, it is possible to use the size of the ionized gas bubble
and infer the time needed for the expansion using an analytical solution such
as the one given by \citet{Spitzer:1978ue} and
\citet{Dyson:1980tp}. This method is commonly used
\citep[e.g.][]{Zavagno:2007ix} and gives a reasonable estimation. 
However, this solution is not exact and assumes a completely
homogeneous medium, therefore the density variations and the turbulence
of the gas are usually neglected. Indeed, \citet{Tremblin:2012he}
showed that the size of the region can be influenced by the turbulence.
In the present paper, we aim at quantifying this effect in order to
build a reliable method that can be used to date OB associations. 

The paper is organized as follows. In Sect.~\ref{sect_simu}, we quantify the
interaction between the ionization of an OB association and the
turbulence of the surrounding molecular gas using 1D and 3D
simulations and comparisons with analytical
solutions. In Sect.~\ref{sect_hrds} we present our investigation of this interaction in
observations by comparing the ionized gas pressure of a sample of
\ion{H}{ii} regions in the HRDS survey \citep{Anderson:2011cu} with
the turbulent ram pressure 
that can be derived from Larson's relations. In Sect.~\ref{sect_age},
we describe how we built a grid of 1D simulations that can be used to
give an estimation of the dynamical age of these regions. The method is
tested on four well-known regions (Rosette, M16, RCW79, and RCW36) for
which we found photometric age estimations in the literature. Finally,
we discuss in Sect.~\ref{sect_disc} the limitations and advantages of
our approach.

\section{Development of \ion{H}{ii} regions in a turbulent medium}\label{sect_simu}

\begin{figure*}[t]
\centering
\includegraphics[width=\linewidth]{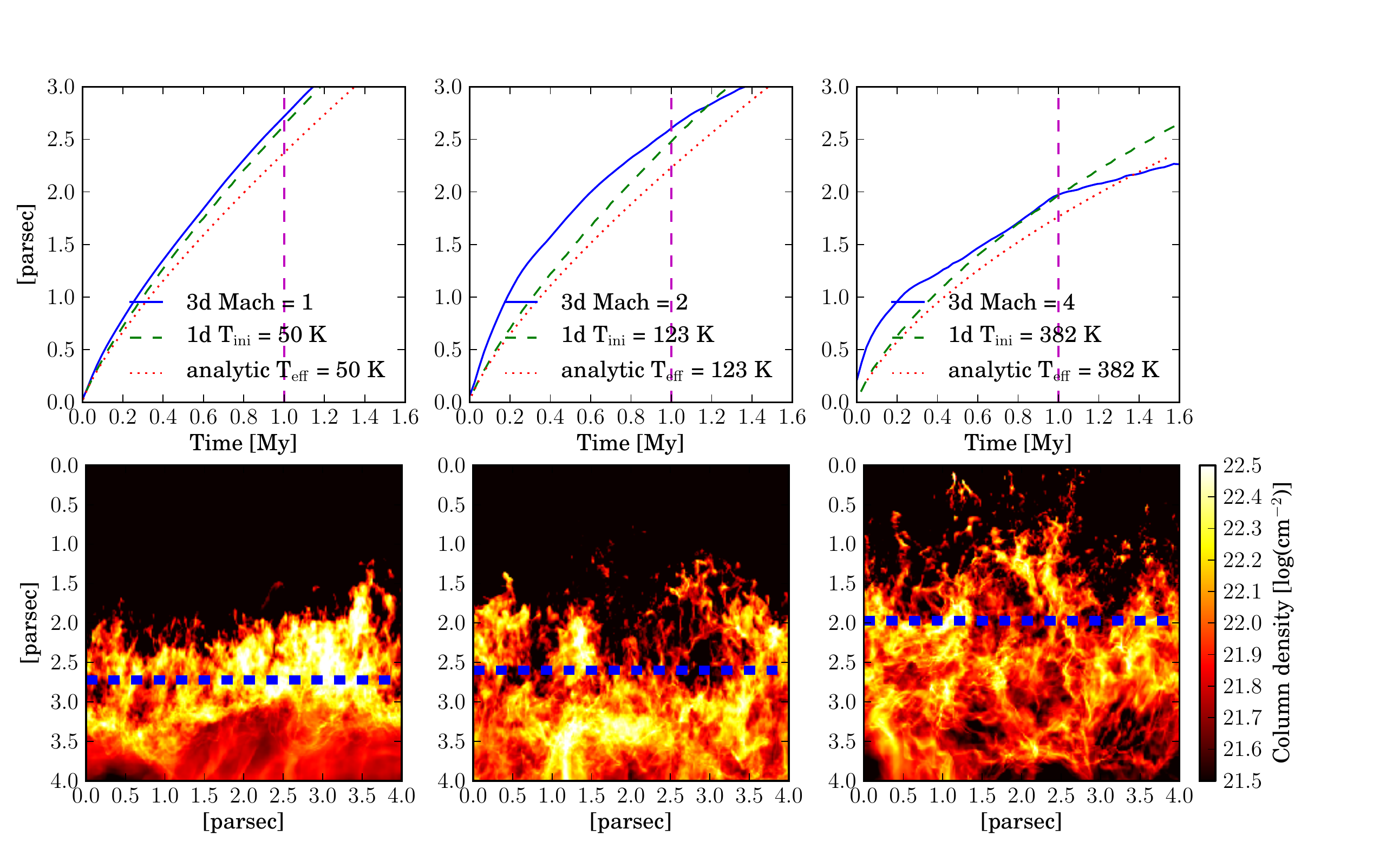}
\caption{\label{fig:simu} Top: Position (relative to the top of the
  box) of the mean ionization front
  as a function of time for the 3D turbulent simulations (respectively
  at Mach 1, 2, and 4 in solid-blue) of
  \citet{Tremblin:2012he} and the corresponding analytical solutions (red-dotted)
  and 1D simulations (green-dashed) in which the turbulence is taking into account by
  an effective temperature/sound speed. Bottom: snapshots of the
  column density of the 3D simulations at 1 Myr after the ionization
  is switched on at the top of the box. The blue-dashed lines show the
mean position of the ionization front.}
\end{figure*}

The expansion of 1D, spherical \ion{H}{ii} regions in a homogeneous
medium is a theoretical exercise whose first solution was proposed by
\citet{Spitzer:1978ue} and \citet{Dyson:1980tp}. Their solution can be written
in terms of the initial Str\"omgren radius $r_s$:
\begin{eqnarray}\label{eq_spitzer}
r_s &=& (3S_*/4\pi n_0^2\alpha)^{1/3}\cr
c_{II}t/r_s &=& 4/7 \times ((r/r_s)^{7/4} -1)\cr
P_{II} &=& n_0 (r_s/r)^{3/2} k_bT_{II}
\end{eqnarray} 
where $S_*$ is the production rate of ionizing photons, $n_0$ the
gas density of the initial homogeneous medium, $\alpha$ the
recombination rate, $c_{II}$ and $T_{II}$ the sound speed and
temperature in the ionized
gas. Although this solution is accurate at early times, it is
obviously wrong for very long times since the radius of the ionized
bubble would diverge to infinity. The radius cannot increase indefinitely
because the expansion is driven by the ionized gas pressure that
decreases like $r^{-3/2}$ (see Eq.~\ref{eq_spitzer}) and the expansion will stop when this
pressure reaches the pressure of the external medium (neglected in the
solution given by Eq.~\ref{eq_spitzer}). A simulation of this
phenomenon can be found in \citet{Tremblin:2011tw} (Fig.~1).

Recently, \citet{Raga:2012fl} proposed a new 1D spherical analytical solution of the
expansion of \ion{H}{ii} regions that takes into account the post-shock
material (and therefore the pressure of the external
medium) in the equation of motion of the ionization front:
\begin{equation}\label{eq_raga}
\frac{1}{c_{II}}\frac{dr}{dt} = \left(\frac{r_s}{r} \right)^{\beta} - \frac{c_0^2}{c_{II}^2}\left(\frac{r}{r_s}\right)^{\beta}
\end{equation}
where $c_0$ is the sound speed in the initial medium, and $\beta$ is
equal to 3/4 for a spherical geometry. Neglecting the
last term in Eq.~\ref{eq_raga} gives back the equation derived by
\citet{Spitzer:1978ue} and \citet{Dyson:1980tp}, for which no equilibrium is
possible. However, equating Eq.~\ref{eq_raga} to zero gives
immediately an equilibrium radius $r_{eq} = r_s (c_{II}/c_0)^{4/3}$
which corresponds to an ionization and hydrostatic equilibrium for
which the pressure of the ionized gas $P_{II}$ is equal to the
pressure of the surrounding medium $P_0$. \citet{Raga:2012fl} compared
this solution to 1D spherical simulations and they both agree
relatively well especially at late times. However, in the simulation, the region
overshoots the equilibrium radius before converging back to it. This
effect is not present in the analytical solution and could be a
consequence of the inertia of the shell that is neglected in the analytical
derivation. 

\begin{figure*}[t]
\centering
\includegraphics[width=\linewidth]{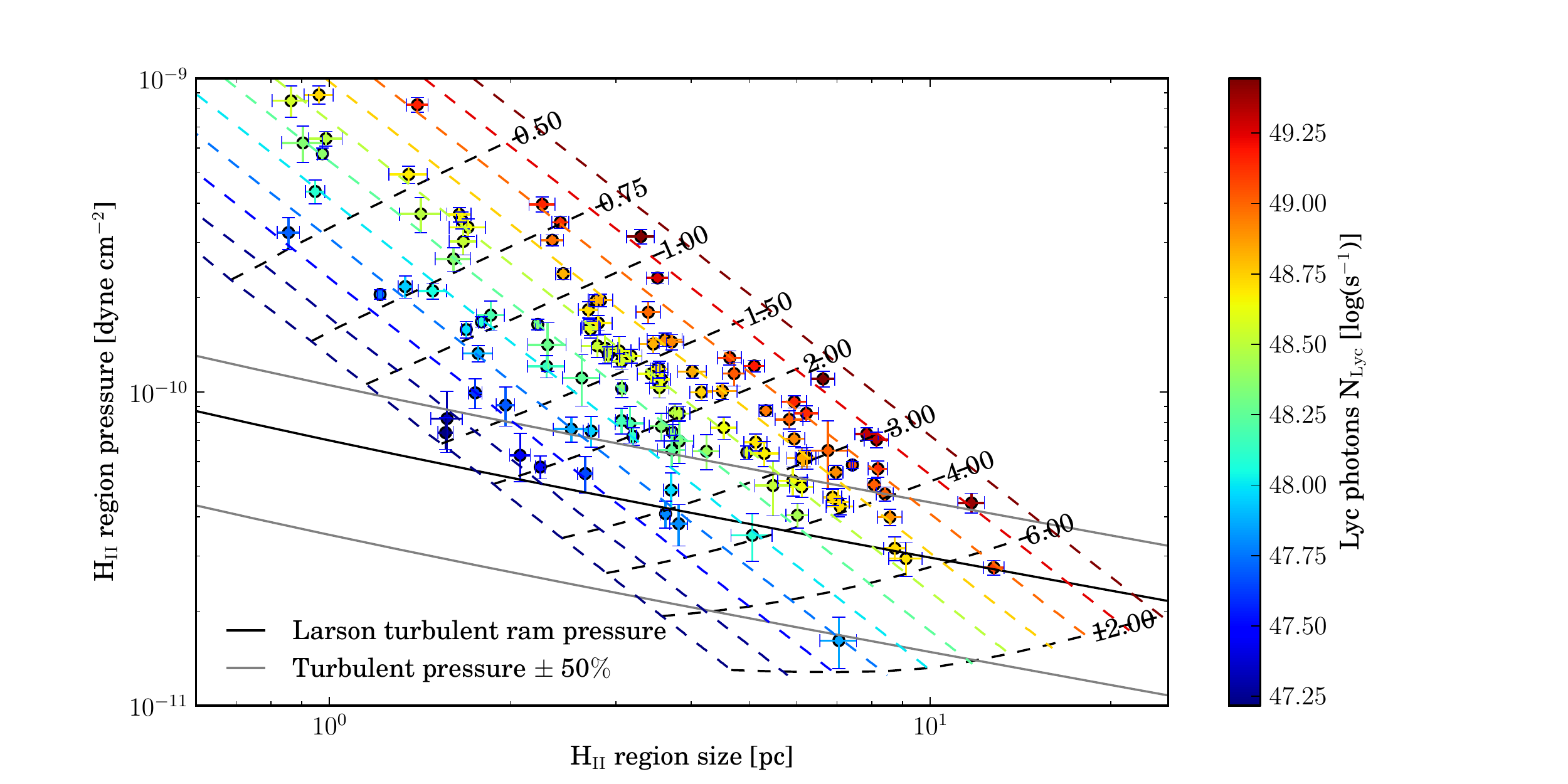}
\caption{\label{fig:ps} Pressure of the \ion{H}{ii} regions of the
  HRDS survey as a function of radius. The black line is the turbulent
ram pressure evaluated from Larson laws, at the scale of the radius of
the region (Eq.~\ref{eq_pturb}). The grey lines show the limits at
$\pm$ 50\% of this relation. The dashed-colored lines are 1D
simulations performed in a density and temperature profile based on
Larson laws at different Lyc fluxes given by the colorbar. The
dashed-black lines are isochrones (in Myr) built from these 
simulations, they are used to estimate the age of the \ion{H}{ii}
regions. To avoid a 3D plot, we assume an constant electron
temperature of 8000 K for all the regions and simulations (only for
this plot).}
\end{figure*}

It has been thought for a long time that this equilibrium cannot be
reached in large and diffuse \ion{H}{ii} regions, because it would take
 much longer than the lifetime of the ionizing sources to reach
it. However, although the thermal
pressure of the initial medium cannot compensate the ionized gas
pressure at early times, \citet{Tremblin:2012he} showed that the
turbulent ram pressure can do the job. We recall this result in
Fig.~\ref{fig:simu}. The bottom panels show snapshots at 1 Myr of the column
density of three different 3D simulations of the ionization of a
turbulent medium respectively with an initial turbulence of Mach 1, 2,
and 4. All the simulations were
performed with
\texttt{HERACLES}\footnote{\url{http://irfu.cea.fr/Projets/Site_heracles}}
\citep{Audit:2011vj}. The box is 4 pc$^3$ at a resolution of 400$^3$, the ionizing
flux at the top of the box is plane-parallel $F_*$ = 10$^{9}$ ph s$^{-1}$ cm$^{-1}$,
the averaged density is $n_0$ = 500 cm$^{-3}$, and there is no gravity  
in these runs. 

The turbulence investigated here is relatively moderate
since one can 
expect a turbulence up to Mach 10 from the Larson's law at this
scale \citep[see][]{Larson:1981vv}. The blue-dashed line in Fig.~\ref{fig:simu}
represents the mean position of the 
ionization front in the simulations and it does show that the
ionization front is slowed down by the supersonic turbulence. In order to
quantify this effect, we adapted Eq.~\ref{eq_raga} to a plane-parallel
geometry ($\beta$ is equal to 1/4) and the equation still
has an analytical solution:
\begin{eqnarray}
r_s &=& F_*/n_0^2\alpha\cr
c_{II}t/r_s &=& f(r/r_s,c_0^2/c_{II}^2)-f(1,c_0^2/c_{II}^2)\cr
f(x,a) &=& 4\tanh^{-1}(\sqrt{a}x^{1/4})/a^{5/2}-4x^{1/4}/a^2-4x^{3/4}/3a
\end{eqnarray} 
To take into account the
turbulence, we replaced the initial
temperature $T_0$ in the sound speed $c_0$ by an effective turbulent temperature:
\begin{equation}
T_{eff} = \langle T_0 + (\mu m_H/k_b) \sigma_0^2 /3 \rangle_\mathrm{box}
\end{equation}
where $\mu$ is the mean molecular weight, $T_0$ and $\sigma_0$ the
temperature and velocity field of the simulation before the ionization
starts, and the rms average is computed on the whole box. A discussion on
the use of an effective turbulent temperature/sound speed can be found
in \citet{MacLow:2004ed}.
Figure~\ref{fig:simu} shows the time evolution of the mean position 
of the ionization front in the 3D simulations, the position given by
the analytical solution and the position of the front in 1D plane-parallel
simulations with the effective turbulent temperature. The analytical
solution and the 1D 
simulations capture quite well the slowing down of the ionization
front caused by the turbulence at late times. At early times, the
ionization front in the 3D simulations propagates faster with
increasing turbulent levels. This is easy to understand: a
  larger turbulence results in denser structures and 
  since the total amount of material is fixed in the box, this implies 
more low density parts. The average
initial Str\"omgren radius computed on the varying density field in
the 3D runs increases with the turbulent level because there are more
and more low density parts for which the Str\"omgren radius will be
large. In the analytical solutions and 1D simulations, the initial
Str\"omgren radius is computed on the average density field, which is
constant at n$_0$ = 500 cm$^{-3}$. Nevertheless, at later times, the
initial conditions do not matter so much anymore, and the 3D
simulations show a slowing down of the propagation of the
  ionization front as the analytical solutions and the 1D
simulations. 

A direct consequence of this analysis is that an \ion{H}{ii} region
will be able to expand in a turbulent medium while $P_{II} \ge
P_\mathrm{turb}$ until the point where the two pressures 
equilibrates. It can be seen with the effective temperatures in
Fig.~\ref{fig:simu} that the turbulent ram pressure can be easily one
order of magnitude bigger than the thermal pressure. Therefore it is
possible that some regions are in equilibrium with their turbulent
environment before the ionizing stars explode.

\begin{figure*}[t]
\centering
\includegraphics[width=\linewidth]{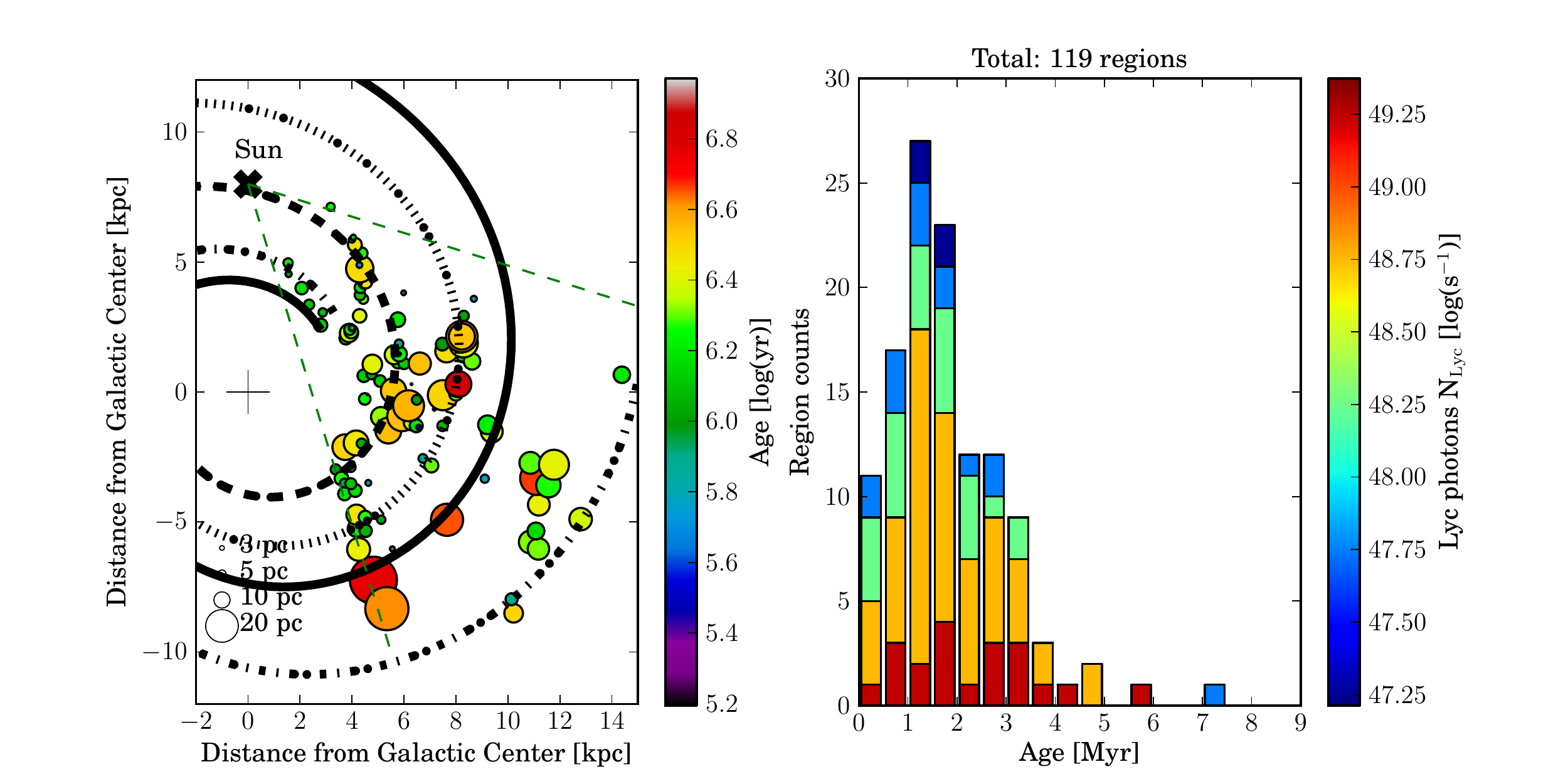}
\caption{\label{fig:sa} Left: Diameter, age, and position of the
  \ion{H}{ii} regions of the HRDS survey in the Galaxy. The green
  dashed lines show the limit of the survey, the black thick curves
  are the position of the spiral arms of our Galaxy based on
  \citet{Russeil:2003ch}. The ages are estimated using the grid of 1D
  simulations at a fixed density at 1pc (3400 cm$^{-3}$) for the
    initial profile. The galactic electron temperature gradient is
    taken into account 
    using Eq.~\ref{eq_te}. Right: age distribution of our sample, the
    ionizing flux of the regions is indicated by the color scale.}
\end{figure*}

\section{HRDS survey and turbulent ram pressure from Larson's laws}\label{sect_hrds}

Using radio continuum surveys of \ion{H}{ii} regions, it is possible
to test that $P_{II} \ge P_\mathrm{turb}$ and to see if the
equilibrium is reached for some regions. We used the \ion{H}{ii}
Region Discovery Survey\footnote{\url{http://www.cv.nrao.edu/hrds}} (HRDS) made with the Green Bank Telescope at 9
GHz whose beam size is 82$''$ \citep[see][]{Anderson:2011cu}. The
ionizing flux $S_*$ and the rms electron density 
$\langle n_{II} \rangle $ in a region can be computed by using
the radio continuum integrated flux
\citep[see][]{MartinHernandez:2005bm}: 
\begin{equation}\label{eq_flux}
\langle n_{II} \rangle = \frac{4.092\times 10^{5}\mathrm{ cm}^{-3}}{\sqrt{b(\nu,T_e)}}
\left(\frac{S_\nu}{\mathrm{Jy}}\right)^{0.5}\left(\frac{T_e}{10^4\mathrm{
    K}}\right)^{0.25}\left(\frac{D}{\mathrm{kpc}}\right)^{-0.5}\left(\frac{\theta_D}{''}\right)^{-1.5}
\end{equation}
\begin{equation}
S_* = \frac{7.603\times 10^{46}\mathrm{ s}^{-1}}{b(\nu,T_e)}\left(\frac{S_\nu}{\mathrm{Jy}}\right)\left(\frac{T_e}{10^4\mathrm{
    K}}\right)^{-0.33}\left(\frac{D}{\mathrm{kpc}}\right)^{2}
\end{equation}
\begin{equation}
b(\nu,T_e) = 1+0.3195\log\left(\frac{T_e}{10^4\mathrm{K}}\right)-0.2130\log\left(\frac{\nu}{\mathrm{GHz}}\right)
\end{equation}
where $S_\nu$ is the radio continuum integrated flux at frequency
$\nu$, $\theta_D$ is the angular diameter of the source, $D$ the
distance from the Sun, and $T_e$ the electron temperature in the
ionized plasma. 

We deconvolved the beam size of the telescope (82$"$) from the
full width at half maximum (FWHM) of the flux given in the HRDS
survey. The HRDS survey provides a complexity flag that indicates
whether the region has a peaked well-defined emission or a complex
multi-component one. We exclude complex regions because the peak of
the emission is not representative of the position of the
ionizing sources, therefore the size of the region is likely to be
wrong. These errors can be corrected with a careful look at the
distribution of the radio emission (e.g. this is done for Rosette and
RCW79 in Sect.~4, their emission has the shape of an annulus) but
cannot be done automatically. This selection gives us a sample of 119
regions for which the 
ionizing flux and the rms electron density can be evaluated. Finally
we converted the FWHM into the 1/e$^2$ width (FWHM$\times$1.7) to have
the angular diameter of the region.

If not measured, $T_e$ can be inferred from the
 galacto-centric distance of the source $R_\mathrm{gal}$:
\begin{equation}\label{eq_te}
T_e = 278 \mathrm{K}\left(\frac{R_\mathrm{gal}}{\mathrm{kpc}}\right) + 6080\mathrm{K}
\end{equation}
This linear relation is an average value of the two samples studied in
\citet{Balser:2011cf} and the average value for the HRDS sample is
8000 K. The ionized gas pressure is then given by:
\begin{equation}\label{eq_pii}
P_{II} = 2\langle n_{II} \rangle k_bT_e
\end{equation}

We estimated the turbulent ram pressure using Larson's laws
\citep[see][]{Larson:1981vv}. Of course the result will be
scale-dependent, and one has to infer which scale is going to matter
when trying to compute this ram pressure. In Sect.~\ref{sect_simu}, we
used the scale of the box to compute the effective sound speed. For
the spherical \ion{H}{ii} regions, we 
make the assumption that the radius of the region is the scale below
which motions will act as an extra pressure against the
expansion. Therefore taking the Larson relations:
\begin{eqnarray}\label{eq_larson}
\langle \sigma \rangle &=& 1.1 \mathrm{km/s} \left(\frac{r}{\mathrm{pc}}\right)^{0.38} \cr
\langle n \rangle &=& 3400 \mathrm{cm}^{-3}
\left(\frac{r}{\mathrm{pc}}\right)^{-1.1} 
\end{eqnarray}
we evaluated the turbulent ram pressure at the scale of the radius of the
region:
\begin{equation}\label{eq_pturb}
P_\mathrm{turb} \approx \langle \rho \rangle (c_0^2+\langle \sigma
\rangle^2/3)
\end{equation}
with $\langle \rho \rangle$ the rms density in g/cm$^3$ and $c_0$ the
sound speed (0.2 km/s). This relation is plotted in Fig.~\ref{fig:ps}
with the ionized gas pressure $P_{II}$ of the regions in the HRDS survey
evaluated from Eq.~\ref{eq_pii}. There is of course a large scatter
around the Larson's relations and uncertainties concerning the right
scale that should be used. Possible variations around this
relations have been investigated in \citet{Lombardi:2010ew} \citep[see
also][]{Hennebelle:2012dk}. Therefore we also plot in
Fig.~\ref{fig:ps} the ram pressure at $\pm$ 50\%. Overall the pressure
of the \ion{H}{ii} regions is greater than the turbulent ram pressure
of the surrounding medium, meaning that they are indeed able to expand in
the turbulent medium. This is consistent with what we infer from the
numerical simulations in Sect.~\ref{sect_simu}. Furthermore, around 20\% of our
sample have pressures at  
$\pm$ 50 \% of the turbulent ram pressure at the scale of their
radius. This suggests that contrary to the usual picture, these
regions may reach the limit of their 
expansion phase and may be in equilibrium with the
surrounding turbulent medium.

\section{Age estimation of the \ion{H}{ii} regions}\label{sect_age}

Based on the comparison in Sect.~\ref{sect_simu} between the 3D turbulent
simulations and the 1D models with an effective turbulent temperature, we
are confident in using 1D spherical simulations as a proxy to evaluate the age
of the \ion{H}{ii} regions. The coupled system of
equation for 1D-spherical hydrodynamics coupled with
ionization/recombination assuming the on-the-spot approximation is
given by
\begin{eqnarray}\label{eq_eulerio}
\frac{\partial \rho} {\partial t}& +&
\frac{1}{r^2}\frac{\partial}{\partial r} \left( r^2 \rho u_r\right) =
0 \quad, \cr 
\frac{\partial n_H X} {\partial t} &+&
\frac{1}{r^2}\frac{\partial}{\partial r} \left( r^2 n_H X u_r\right) =
 n_H(1-X)\sigma_\gamma F_\gamma  -\alpha X^2 n_H^2 \quad, \cr
\frac{\partial \rho u_r}{\partial t}& +&
\frac{1}{r^2}\frac{\partial}{\partial r} \left(  r^2 \rho u_r^2\right)
 =  - \frac{\partial p}{\partial r} \quad, \cr 
\frac{\partial E} {\partial t} &+&
\frac{1}{r^2}\frac{\partial}{\partial r} \left( r^2(E + p)u_r\right)
  =  n_H(1-X)\sigma_\gamma F_\gamma e_\gamma \cr
&& \qquad \qquad \qquad \qquad -\alpha X^2 n_H^2 k_b T/(\gamma-1)\quad, \cr
\frac{1}{r}\frac{\partial}{\partial r} \left(r F_\gamma \right)&=& n_H (1-X)\sigma_\gamma
(-F_\gamma + F_*\delta_{0}(r-r_*))
\end {eqnarray}
where $\rho$, $u_r$, P, T, are the density, velocity,
  pressure, and temperature of
the fluid, $E$ is the total energy given by $E=\rho u_r^2/2+e$, $X$ the ionization fraction given by $n_{H^+}/n_H$, $n_H$ the total hydrogen
density $n_{H^+} + n_{H^0}$, $F_\gamma$ the ionizing flux,
$\sigma_\gamma$ the ionization cross section, $e_\gamma$ the mean energy
given by an ionizing photon to the gas, $\alpha$ the recombination
rate, and $\gamma$ the adiabatic index. We assume an ideal gas to link
the internal energy $e$ to the pressure: $p=e(\gamma-1)$. The system is splitted in
a hydrodynamic step solved explicitly by a Godunov exact solver and an
ionization/recombination step solved implicitly. We used a $\gamma$ of
1.001 so that the gas is locally isothermal in the absence of
ionization/recombination processes.  

We performed 1D spherical models
in a medium with a density and pressure profile given by
Eq.~\ref{eq_larson} and Eq.~\ref{eq_pturb}. When the ionized bubble
expands in such a density/pressure profile, it will ``feel'' the
turbulent ram pressure at a given radius acting against the
expansion. A grid of models was computed with different fluxes
(log$_{10}$(S$_*$) between 47. and 51. in steps of 0.25),
electronic temperature (T$_e$: 10$^4$ K $\pm$ 50\% in steps of
1000 K), and density at 1 pc ($n$(1pc): 3400 cm$^{-3}$ $\pm$ 50\% in steps of
340 cm$^{-3}$). The simulation domain extends up to 25 parsec, we took
a resolution of 2500 cells (uniformly spaced) and ran the
simulations during 12 Myr. These models are plotted in
Fig.~\ref{fig:ps} for a fixed electron temperature 
$T_e$ = 8000 K and a density at 1 pc of $n$(1pc) = 3400 cm$^{-3}$. In
our final age estimation, we do take into account the electron
temperature using Eq.~\ref{eq_te}, we used a fixed temperature in
Fig.~\ref{fig:ps} to avoid a 3D plot for our grid of simulations,
however we do take into account the electron temperature
dependence for our final age estimation. 
The regions from the HRDS survey fall exactly on the simulated tracks
at their ionizing fluxes. This is normal and a consequence of photon
conservation that is assumed in Eq.~\ref{eq_flux} to evaluate $S_*$
from $\langle n_{II} \rangle$ and is also assumed in the
simulations. It can be shown using photon conservation that $P_{II}
\propto S_*^{0.5}r^{-1.5}$ (see Eq.\ref{eq_spitzer}), which gives the linear tracks in log space
for Fig.~\ref{eq_flux}. The real contribution of the simulations are
the isochrone curves built out of them (black-dashed
lines). We can then estimate the dynamical age of the different
regions. 

\begin{table}  
\begin{center}  
\begin{tabular}{lrrrr}  
\hline \hline   
Cloud ($D$) & Radius & $S_\nu(\nu) $ & Phot. Age & Dyn. Age \\

[kpc]   &[pc]& [Jy](GHz)     & [Myr] & [Myr]  \\ 
\hline       
Rosette (1.6$^a$)  & 18.7$\pm$1.2$^b$ & 350(4.75)$^b$ &  $\le$ 5$^c$    & 5.0$\pm$0.4 \\
M16 (1.75$^d$) &  7.2$\pm$0.7$^e$ & 117(5)$^e$    &  2-3$^f$ & 1.9$\pm$0.2 \\
RCW79 (4.3$^g$)  &  7.1$\pm$0.3$^h$ & 19.5(0.84)$^h$ &  2-2.5$^i$       & 2.2$\pm$0.1 \\
RCW36 (0.7$^j$)  &  1.1$\pm$0.07$^e$ & 30(5)$^e$     &  1.1$\pm$0.6$^k$ & 0.4$\pm$0.03 \\
\hline   
\end{tabular}  
\end{center}  
\caption{\ion{H}{ii} regions used for comparison between photometric
  and dynamical age. The fluxes are integrated radio continuum fluxes
  at the frequency indicated in parenthesis. The electron Temperature
  $T_e$ is determined from 
Eq.~\ref{eq_te}. $^a$\citep{RomanZuniga:2008tm}
$^b$\citep{Celnik:1985th} $^c$\citep{Martins:2012gv}
$^d$\citep{Guarcello:2007im} $^e$ \citep{Condon:1993kf}
$^f$\citep{Hillenbrand:1993gu} $^g$\citep{Russeil:2003ch} 
$^h$ \citep{Mauch:2003eda} $^i$\citep{Martins:2010jw} 
$^j$\citep{Yamaguchi:1999ww} $^k$\citep{Ellerbroek:2013cz}\label{tab_hii}}
\end{table} 

We tested this method on well-known regions for which an independent age of the
central massive OB stars is available from photometry and
evolutionary tracks in the HR diagram. We took four
regions: Rosette, M16, RCW79, 
and RCW36, their parameters and the corresponding references are given
in Tab.~\ref{tab_hii}. For all the regions, we used Eq.~\ref{eq_te} to
estimate the electron temperature. Our age estimations are in good
agreement with the ages derived from photometry, however some
questions can be raised:
\begin{itemize}
\item
In the case of the Rosette Nebula,
\citet{Martins:2012gv} concluded that the age of the two most massive O
stars in NGC2244 is less than 2 Myr, however they suggest that either
there is a bias in their effective temperature, or they were the
last to form. For the second possibility, the \ion{H}{ii} region would
be powered first by the lower-mass O stars, and then by
the two most massive ones that dominate the total ionizing flux. The
ionizing flux is thus a function of time in that case, and could
change the dynamical age. The same may also happen for the future
development of RCW79 with the appearance of 
a compact and younger \ion{H}{ii} region at the southeast of
the region.
\item
The dynamical age of RCW36 is at the lower end of the photometric
range. This could mean that the ionizing O star is also the last to
form. However contrary to the Rosette Nebula, we do not expect a
two-stage expansion in this case because there is only one dominant
ionizing source. Other physical phenomena at the early stages of the
expansion could also delay the expansion by 0.1-0.2 Myr.
(see Sect.~5.1).
\end{itemize}

In Sect.~5.2, we will discuss in further detail the
possible biases of our 
approach. However, considering the uncertainties and 
the error bars of the age derived from evolutionary stages, the
dynamical age agrees well. Besides,
it has the advantage of being much simpler to compute and can be
applied to a large sample of OB associations. We apply the method on the
HRDS sample and give the age distribution in the right panel of
Fig.~\ref{fig:sa}. The average age of the 119 regions is 1.9 Myr $\pm$
0.7 Myr. The distribution is plotted as a stacked histogram indicating
the ionizing flux of the sources and we do not see any trend as a
function of flux. There is a tail with older regions that is not very well
sampled, probably because there are not many
big and bright sources such as the Rosette Nebula in the survey. The
average age of the sample is relatively small compared to the typical
lifetime of a massive star (30 Myr for a typical B-type star
  of 10 M$_\odot$). 
The left panel of Fig.~\ref{fig:sa} shows the size, age and position of
the regions in the Galaxy. The Scutum-Crux arm (dash-dotted spiral)
has been extended beyond the distance constrained in
\citet{Russeil:2003ch}, this is probably why the most distant regions
do not lie very well on the arm.  At first sight, the regions in the 
Sagittarius-Carina arm (dashed spiral) seem younger than the regions
in the other arms, however most of these regions are also the closest
to the Sun so the completeness and sensitivity limits of the survey have to be
carefully studied before making any firm conclusion. 

\begin{figure}[t]
\centering
\includegraphics[width=\linewidth]{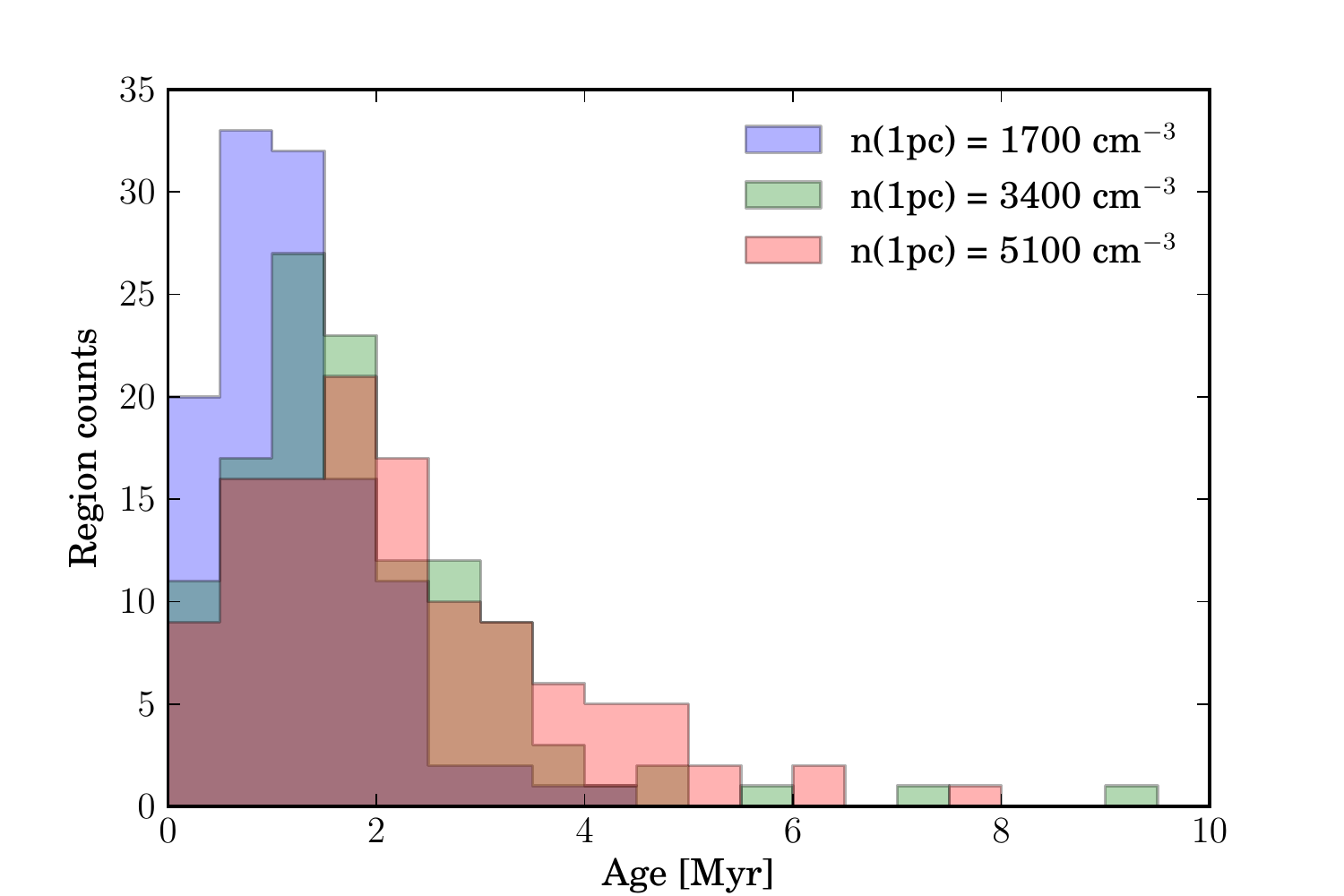}
\caption{\label{fig:histrho} Effect of changing by $\pm$ 50\% the density at 1 pc of
  the initial density profile on the total age distribution. Note that this
  effect is applied to all the regions, in reality it should not be
  such a systematic positive or negative bias.}
\end{figure}

\section{Discussion}\label{sect_disc}

\subsection{What about gravity and magnetic fields ?}

Other physical phenomena could also affect the expansion of the
ionization front and balance
the ionized gas pressure. Their associated pressure should be compared
to the turbulent pressure given by Eq.~\ref{eq_pturb} and with the
ionized-gas pressure in Eq.~\ref{eq_spitzer}. The effect of self gravity is negligible on
the expansion because of Gauss's law for gravity. Only the mass of the
ionized gas will act at the ionization front (however the gas self gravity
is an important 
effect to consider when studying the compression in the shell). The gravity
of the central cluster could play a role. If we assume that an
hydrostatic equilibrium is possible, we have the relation:
\begin{eqnarray}
 P_{II}-P_0 &= &\int^\infty_r \frac{GM\rho(r')}{r'^2} dr' \quad
 \mathrm{with} \quad \rho(r') = \rho(1\mathrm{pc})\left(\frac{r'}{\mathrm{pc}}\right)^{-1.1} \cr
P_{II} &\approx& P_G\approx\frac{GM\rho(1\mathrm{pc})}{2.1\times(1 \mathrm{pc})} \left(\frac{r}{\mathrm{pc}}\right)^{-2.1} 
\end{eqnarray}
where $M$ is the mass of the central cluster. Assuming a cluster mass
of 500 M$_\odot$ gives a gravitational pressure on the order of
1.62$\times$10$^{-10}$ dyne cm$^{-2}$ at one parsec, which is comparable to the
turbulent pressure at that scale (see
Fig.~\ref{fig:ps}). Nevertheless, this pressure is dropping 
much faster than the turbulent pressure ($\approx$ r$^{-2}$),
therefore if expansion has happened at some point with $P_{II}$ $\ge$ $P_G$,
the gravity of the central cluster will never be able to stop the
large-scale expansion in the future. However, gravitational effects
can be important at 
small scales for compact and ultra-compact \ion{H}{ii} regions
\citep[see][]{Keto:2007jy,Peters:2010bo}, therefore we do not expect our model to
apply for spatial scale smaller than $\approx$ 0.1 pc (and age smaller
than 0.1-0.2 Myr).

The magnetic pressure could also play a role. Assuming a constant mass
to flux ratio in molecular clouds and a typical magnetic field of 20
$\mu G$ at one parsec \citep[see][]{Troland:2008kp,Lazarian:2012jh},
we can derive the scaling relation (for larger scales): 
\begin{eqnarray}
\frac{\langle \phi \rangle}{\phi(1\mathrm{pc})}&=& \frac{\langle M
  \rangle}{M(1\mathrm{pc})} \rightarrow \frac{\langle B \rangle}{B(1\mathrm{pc})}\times
\left(\frac{r}{\mathrm{pc}} \right)^2= \frac{\langle \rho
  \rangle}{\rho(1\mathrm{pc})}\times \left(\frac{r}{\mathrm{pc}}
\right)^3\cr 
\langle B \rangle &=& 20 \mu G
\left(\frac{r}{\mathrm{pc}}\right)^{-0.1} \quad \mathrm{and} \quad P_B
= \frac{\langle B\rangle^2}{8\pi} 
\end{eqnarray}

At one parsec, this magnetic field leads to a magnetic pressure on the
order of 1.5$\times$10$^{-11}$ dyne cm $^{-2}$, much smaller than the
turbulent pressure. Even if the magnetic pressure inside the
  \ion{H}{ii} region acts as a support, its strength is small compared
to the ionized-gas pressure. Therefore the magnetic field pressure does not
have an important effect for the large scale evolution of \ion{H}{ii}
regions. However, \citet{Crutcher:2012hw} showed that the magnetic
field can be on the order of 100-1000 $\mu G$ for dense gas at small
scales, consequently the small-scale evolution of \ion{H}{ii} regions
could depend on the magnetic field. 

Both gravity and magnetic fields should be considered for the
evolution of compact and ultra-compact \ion{H}{ii} regions but should
not affect 
the large-scale evolution of diffuse nebulae. The presence of dust
can also have an important effect for small regions
\citep[see][]{Inoue:2002ht,Arthur:2004id}. 
As a consequence, we do
not expect our models to be able to predict the ages for small regions
at a scale of 0.1 pc and the error bars for regions around 1 pc are
possibly larger because of the uncertainties in the small scale
evolution. 

\subsection{The effect of initial conditions}

Environmental variations in the surrounding density can also be an
issue. We managed to correct for the electron temperature but we
did not take into account possible variations of the density profile
around the one given by Eq.~\ref{eq_larson}. Our grid of simulations
can take into account such variations (up to $\pm$ 50\% around $n$(1pc) = 3400
cm$^{-3}$) but we do not have observational constraints for the
local density around the regions of the HRDS survey. To illustrate the
effect of the density variations, we recomputed the age distribution
of Fig.~\ref{fig:sa} for $n$(1pc) = 1700 cm$^{-3}$ and 5100
cm$^{-3}$. The corresponding distributions are given in
Fig.~\ref{fig:histrho}. For $n$(1pc) = 1700 cm$^{-3}$, the distribution
is shifted at 1.4 Myr $\pm$ 0.8 Myr, and for  $n$(1pc) = 5100 cm$^{-3}$
at 2.3 Myr $\pm$ 0.8 Myr. The changes are significant, but we do not
expect the local variation in the density at one parsec to be
systematically positive or 
negative for all regions so that we would have to shift the
  global Larson's law accordingly. Therefore the average age of the
distribution should remain 
relatively constant and the variations around the density at one
parsec may only increase the standard deviation. 
Furthermore, we do not expect the bias caused by the density
variations to be important for the large regions. Indeed, large
regions have a surrounding area big enough to recover an average 
density that should be relatively close to what can be inferred from
 Larson's relation. For smaller regions ($\approx$ 1 pc), the local density variations
can be important and lead to different ages. However, if enough
regions are included in the statistics, the age distribution should
be fairly good although the age estimation of a small particular region might be
wrong.

\subsection{Advantages of dynamical age determinations and perspectives}

The dynamical evolution inferred
from our grid of models is a good way to get an estimation of
the age of the OB associations, especially when we consider how
difficult and uncertain it is to get it from photometry and
evolutionary tracks. In principle 3D simulations would be required to
study the evolution in a turbulent medium but, they are currently too
time-consuming to allow the computation of a full grid of
models as a function of flux, temperature and density. Thanks to the
equivalent grid of 1D simulations, the method is relatively cheap and
we can take into account the environmental dependences. 

Although this method cannot be used for small regions for which
magnetic fields and gravity have to be considered, we can
easily date a large sample of diffuse
\ion{H}{ii} regions when they can be resolved and their Lyc flux
estimated in observations. This method could also be applied to
nearby extra-galactic \ion{H}{ii} regions, thus allowing us to get age
distributions of massive-star forming regions in other galaxies. These
distributions could then be used to constrain galaxy-scale simulations of
star formation.

%
%

\begin{acknowledgements}
We would like to
acknowledge the Nordita program on Photo-Evaporation in Astrophysical
Systems (June 2013) where part of the work for this paper was carried
out. We also thank L. Deharveng, D. Russeil, J. Tig\'e, G. Chabrier, and
E. Audit for valuable discussions. N.S. acknowledges support by the
ANR-11-BS56-010 project ``STARFICH''. This work is partly supported by
the European Research Council under the European Community's Seventh
Framework Programme (FP7/2007−2013 Grant Agreement No. 247060).
\end{acknowledgements}

%
%


\bibliographystyle{aa}
\bibliography{main.bib}

\begin{thebibliography}{33}
\expandafter\ifx\csname natexlab\endcsname\relax\def\natexlab#1{#1}\fi

\bibitem[{Anderson {et~al.}(2011)Anderson, Bania, Balser, \&
  Rood}]{Anderson:2011cu}
Anderson, L.~D., Bania, T.~M., Balser, D.~S., \& Rood, R.~T. 2011, ApJ, 194, 32

\bibitem[{Arthur {et~al.}(2004)Arthur, Kurtz, Franco, \&
  Albarr{\'a}n}]{Arthur:2004id}
Arthur, S.~J., Kurtz, S.~E., Franco, J., \& Albarr{\'a}n, M.~Y. 2004, ApJ, 608,
  282

\bibitem[{Audit {et~al.}(2011)Audit, Gonz{\'a}lez, Vaytet, Fromang, Hennebelle,
  Teyssier, Tremblin, \& Thooris}]{Audit:2011vj}
Audit, E., Gonz{\'a}lez, M., Vaytet, N., {et~al.} 2011, Astrophysics Source
  Code Library, 02016

\bibitem[{Balser {et~al.}(2011)Balser, Rood, Bania, \&
  Anderson}]{Balser:2011cf}
Balser, D.~S., Rood, R.~T., Bania, T.~M., \& Anderson, L.~D. 2011, ApJ, 738, 27

\bibitem[{Celnik(1985)}]{Celnik:1985th}
Celnik, W.~E. 1985, A{\&}A, 144, 171

\bibitem[{Condon {et~al.}(1993)Condon, Griffith, \& Wright}]{Condon:1993kf}
Condon, J.~J., Griffith, M.~R., \& Wright, A.~E. 1993, AJ, 106, 1095

\bibitem[{Crutcher(2012)}]{Crutcher:2012hw}
Crutcher, R.~M. 2012, ARA\&A, 50, 29

\bibitem[{Dyson \& Williams(1980)}]{Dyson:1980tp}
Dyson, J.~E. \& Williams, D.~A. 1980, {Physics of the interstellar medium}
  (Manchester University Press)

\bibitem[{Ellerbroek {et~al.}(2013)Ellerbroek, Bik, Kaper, Maaskant, Paalvast,
  Tramper, Sana, Waters, \& Balog}]{Ellerbroek:2013cz}
Ellerbroek, L.~E., Bik, A., Kaper, L., {et~al.} 2013, A{\&}A, 558, 102

\bibitem[{Guarcello {et~al.}(2007)Guarcello, Prisinzano, Micela, Damiani,
  Peres, \& Sciortino}]{Guarcello:2007im}
Guarcello, M.~G., Prisinzano, L., Micela, G., {et~al.} 2007, A{\&}A, 462, 245

\bibitem[{Hennebelle \& Falgarone(2012)}]{Hennebelle:2012dk}
Hennebelle, P. \& Falgarone, E. 2012, A\&AR, 20, 55

\bibitem[{Hillenbrand {et~al.}(1993)Hillenbrand, Massey, Strom, \&
  Merrill}]{Hillenbrand:1993gu}
Hillenbrand, L.~A., Massey, P., Strom, S.~E., \& Merrill, K.~M. 1993, AJ, 106,
  1906

\bibitem[{Inoue(2002)}]{Inoue:2002ht}
Inoue, A.~K. 2002, AJ, 570, 688

\bibitem[{Keto(2007)}]{Keto:2007jy}
Keto, E. 2007, ApJ, 666, 976

\bibitem[{Larson(1981)}]{Larson:1981vv}
Larson, R.~B. 1981, MNRAS, 194, 809

\bibitem[{Lazarian {et~al.}(2012)Lazarian, Esquivel, \&
  Crutcher}]{Lazarian:2012jh}
Lazarian, A., Esquivel, A., \& Crutcher, R. 2012, ApJ, 757, 154

\bibitem[{Lombardi {et~al.}(2010)Lombardi, Alves, \& Lada}]{Lombardi:2010ew}
Lombardi, M., Alves, J., \& Lada, C.~J. 2010, A{\&}A, 519, L7

\bibitem[{Mac~Low \& Klessen(2004)}]{MacLow:2004ed}
Mac~Low, M.-M. \& Klessen, R.~S. 2004, Reviews of Modern Physics, 76, 125

\bibitem[{Mart{\'\i}n-Hern{\'a}ndez {et~al.}(2005)Mart{\'\i}n-Hern{\'a}ndez,
  Vermeij, \& van~der Hulst}]{MartinHernandez:2005bm}
Mart{\'\i}n-Hern{\'a}ndez, N.~L., Vermeij, R., \& van~der Hulst, J.~M. 2005,
  A{\&}A, 433, 205

\bibitem[{Martins {et~al.}(2012)Martins, Mahy, Hillier, \&
  Rauw}]{Martins:2012gv}
Martins, F., Mahy, L., Hillier, D.~J., \& Rauw, G. 2012, A{\&}A, 538, 39

\bibitem[{Martins {et~al.}(2010)Martins, Pomar{\`e}s, Deharveng, Zavagno, \&
  Bouret}]{Martins:2010jw}
Martins, F., Pomar{\`e}s, M., Deharveng, L., Zavagno, A., \& Bouret, J.~C.
  2010, A{\&}A, 510, 32

\bibitem[{Mauch {et~al.}(2003)Mauch, Murphy, Buttery, Curran, Hunstead,
  Piestrzynski, Robertson, \& Sadler}]{Mauch:2003eda}
Mauch, T., Murphy, T., Buttery, H.~J., {et~al.} 2003, MNRAS, 342, 1117

\bibitem[{Meynet \& Maeder(2003)}]{Meynet:2003kc}
Meynet, G. \& Maeder, A. 2003, A\&A, 404, 975

\bibitem[{Peters {et~al.}(2010)Peters, Mac~Low, Banerjee, Klessen, \&
  Dullemond}]{Peters:2010bo}
Peters, T., Mac~Low, M.-M., Banerjee, R., Klessen, R.~S., \& Dullemond, C.~P.
  2010, ApJ, 719, 831

\bibitem[{Raga {et~al.}(2012)Raga, Canto, \& Rodr{\'\i}guez}]{Raga:2012fl}
Raga, A.~C., Canto, J., \& Rodr{\'\i}guez, L.~F. 2012, MNRAS, 419, L39

\bibitem[{Rom{\'a}n-Z{\'u}{\~n}iga \& Lada(2008)}]{RomanZuniga:2008tm}
Rom{\'a}n-Z{\'u}{\~n}iga, C.~G. \& Lada, E.~A. 2008, Handbook of Star Forming
  Regions, I, 928

\bibitem[{Russeil(2003)}]{Russeil:2003ch}
Russeil, D. 2003, A{\&}A, 397, 133

\bibitem[{Spitzer(1978)}]{Spitzer:1978ue}
Spitzer, L. 1978, {Physical processes in the interstellar medium} (New York
  Wiley-Interscience)

\bibitem[{Tremblin {et~al.}(2012)Tremblin, Audit, Minier, Schmidt, \&
  Schneider}]{Tremblin:2012he}
Tremblin, P., Audit, E., Minier, V., Schmidt, W., \& Schneider, N. 2012,
  A{\&}A, 546, 33

\bibitem[{Tremblin {et~al.}(2011)Tremblin, Audit, Minier, \&
  Schneider}]{Tremblin:2011tw}
Tremblin, P., Audit, E., Minier, V., \& Schneider, N. 2011, in Astronum 2010,
  San Diego, 87

\bibitem[{Troland \& Crutcher(2008)}]{Troland:2008kp}
Troland, T.~H. \& Crutcher, R.~M. 2008, ApJ, 680, 457

\bibitem[{Yamaguchi {et~al.}(1999)Yamaguchi, Mizuno, Saito, Matsunaga, Mizuno,
  Ogawa, \& Fukui}]{Yamaguchi:1999ww}
Yamaguchi, N., Mizuno, N., Saito, H., {et~al.} 1999, PASJ, 51, 775

\bibitem[{Zavagno {et~al.}(2007)Zavagno, Pomar{\`e}s, Deharveng, Hosokawa,
  Russeil, \& Caplan}]{Zavagno:2007ix}
Zavagno, A., Pomar{\`e}s, M., Deharveng, L., {et~al.} 2007, A{\&}A, 472, 835

\end{thebibliography}

\end {document}